\documentclass[runningheads]{llncs}
\usepackage{graphicx}
\usepackage{braket}
\usepackage{physics}
\usepackage{amssymb,amsfonts,amsmath}

  \newcommand{\eq}{Eq.~}
\newcommand{\eqs}{Eqs.~}
\newcommand{\fig}{Fig.~}

\newcommand{\cf} {cf.~}
\newcommand{\ug} {\!=\!}

\newcommand{\eg} {e.g.~}

\newcommand{\rref} {Ref.~}

\begin{document}
\title{Quantum non-Markovian collision models \\from colored-noise baths}
%
%
\author{Dario Cilluffo\inst{1,2} and Francesco Ciccarello\inst{1,2}}

\authorrunning{D. Cilluffo and F. Ciccarello}
%
\institute{Universit\`a degli Studi di Palermo, Dipartimento di Fisica e Chimica -- Emilio Segr\'e, via Archirafi 36, I-90123 Palermo, Italia \and
NEST, Istituto Nanoscienze-CNR, Piazza S. Silvestro 12, 56127 Pisa, Italy\\
\email{francesco.ciccarello@unipa.it}}
\maketitle            
\begin{abstract}
	A quantum collision model (CM), also known as repeated
	interactions model, can be built from the standard microscopic
	framework where a system $S$ is coupled to a white-noise bosonic bath
	under the rotating wave approximation, which typically results in Markovian dynamics. Here, we discuss how to generalize the CM construction to
	the case of frequency-dependent system-bath coupling, which defines a class of colored-noise baths. This leads to an intrinsically non-Markovian CM, where each ancilla (bath subunit) collides repeatedly with $S$ at different steps. We discuss the illustrative example of an atom in front a mirror in the regime of non-negligible retardation times. 
\end{abstract}



%
%
%
\section{Introduction}\label{intro}

Dynamics of open quantum systems \cite{books1,books2} currently plays a central role in a number of research areas concerned in various ways with quantum coherence effects. A theoretical tool to tackle open dynamics is embodied by quantum collision models (CMs) \cite{rau,scarani2002,brun,review-CM}. Recent times have seen growing use of CMs in fields such as quantum thermodynamics \cite{esposito}, where most frequently they go under the name of {\it repeated interactions models}, quantum non-Markovian dynamics \cite{review-vega}, quantum gravity \cite{altamirano2017a} and quantum optics \cite{grimsmo2015,whalen2017,pichler,vuckovic,milburn2017,fischerJPC}. While most treatments in fact assume that the system-environment dynamics is at the microscopic level described by a CM, there exist scenarios where a CM arises instead as an effective descriptive {\it picture} from a standard microscopic model of a bosonic bath coupled to the open system $S$ under the rotating wave approximation \cite{CMQO,vuckovic,milburn2017}. The construction of such CMs, discussed by one of the authors in a recent paper \cite{CMQO}, relies crucially on having a {\it white}-noise bath, which usually means that both the reservoir density of states and coupling strength between $S$ and bath normal modes are independent of frequency. In typical cases, this assumption typically leads to Markovian dynamics. This is reflected by the memoryless nature of the associated CM, in particular the fact that $S$ collides with one bath subunit at a time and that each given subunit (also called ``ancilla") collides with the system only once [see \fig1(a)].
\begin{figure}
	\includegraphics[width=\textwidth]{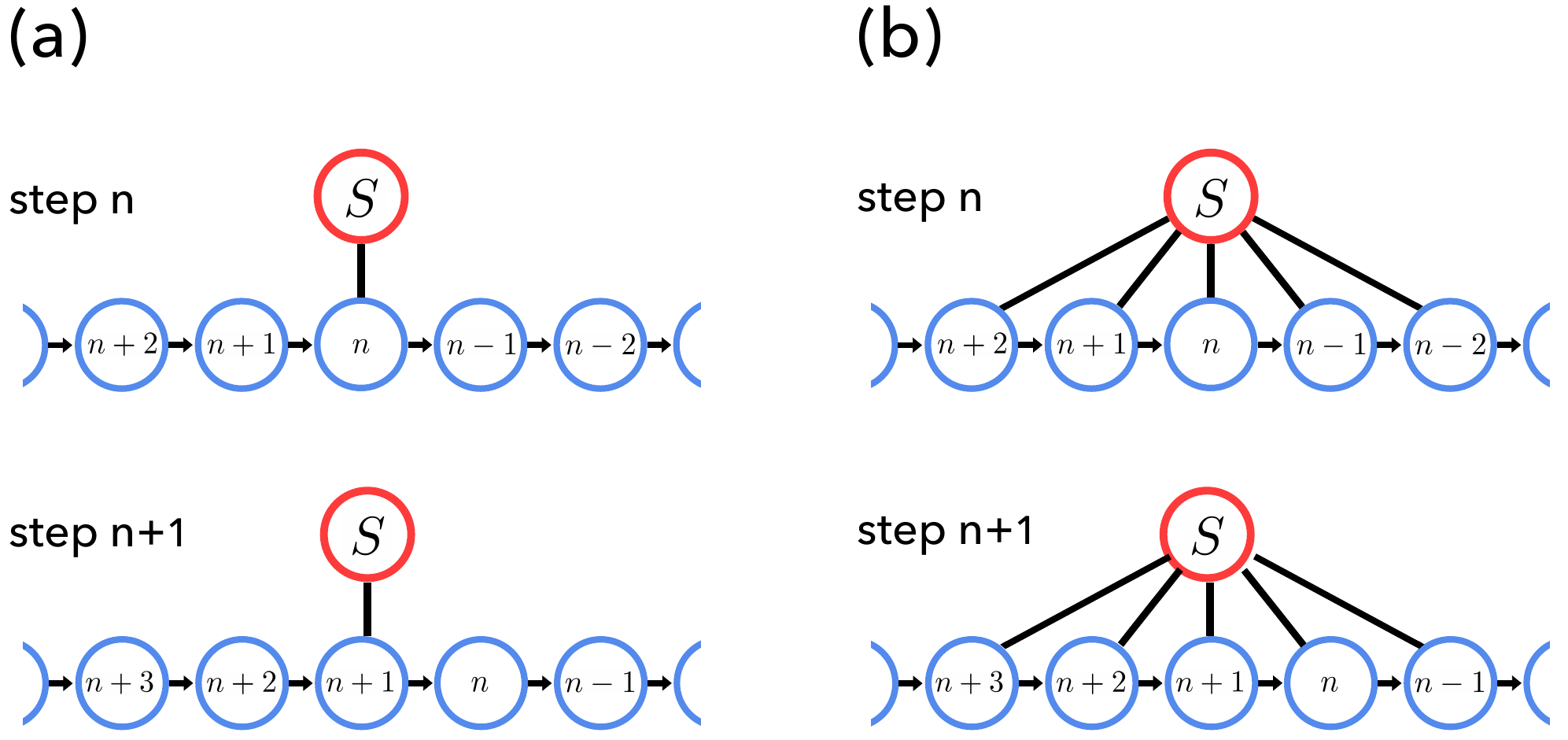}
	\caption{(a) Sketch of the memoryless collision model arising for white coupling: $S$ collides in succession with bath ancillas, one at a time. Each given ancilla collides with $S$ only once. Note that we use a conveyor-belt representation, where ancillas move to the right and are labeled by an index that grows from right to left. (b) Pictorial representation of the general non-Markovian collision model arising from a bath with colored couplings. At each step, $S$ generally interacts with all the ancillas, not only one as in (a). Thereby, each given ancilla collides with $S$ repeatedly.} \label{fig1}
\end{figure}

During discussions at the 684th WE-Heraus-Seminar, the question was raised as to whether a CM can still be worked out when the white-noise assumption is relaxed and, if so, what are its main features. In this short paper we address this issue, which extends the discussion developed in \rref\cite{CMQO}, in the case of a colored system-bath coupling and show that a CM can still be constructed. This however differs from the basic memoryless version of a CM in that $S$ generally collides with many ancillas at a time in such a way that each given ancilla undergoes multiple collisions at different steps with the system [see \fig1(b)]. These features make the resulting CM intrinsically non-Markovian, in that the open dynamics of $S$ is not decomposable into a sequence of elementary completely positive maps and thus cannot be described by a Lindblad master equation in the continuous-time limit.

This paper is organized as follow. We start in Section \ref{defCM} with a concise review of CMs. In Section \ref{bos-rev}, we reconsider the model of \rref\cite{CMQO}, featuring a system in contact with a bosonic reservoir \cite{CMQO}, but generalized to the case of coloured couplings. In Section \ref{tm}, we present the description of the reservoir in terms of time modes, which is then used to construct the CM in Section \ref{timedis}. In Section \ref{mirror}, we discuss the important example of an atom coupled to a semi-infinite waveguide in the regime of non-negligible time delays, showing that it gives rise to a CM where each ancilla undergoes two delayed collisions with the system. This non-Markovian CM is used in Section \ref{DDE} to retrieve the known (time-non-local) delay differential equation governing atomic decay in the setup of Section \ref{mirror}. Finally, we draw our conclusions in Section \ref{concl}.

\section{Brief review of memoryless collision models}\label{defCM}

The standard formulation of a CM imagines a quantum system $S$ (the open system) interacting with a quantum bath, the latter being comprised of a large number of identical non-interacting subunits, the ``ancillas" [see \fig1(a)]. The joint system ($S$ plus bath) starts in the factorized state $\sigma_0=\rho_{0}\otimes\left(\eta\otimes\eta\otimes...\right)$ with $\rho_0$ ($\eta$) the initial state of $S$ (each ancilla). In the remainder, tensor product symbols will be omitted. 
As sketched in \fig1(a), the dynamics occurs through sequential, pairwise, short interactions (collisions) between $S$ and each ancilla: $S$-(ancilla 1), $S$-(ancilla 2),...\,. Importantly, before colliding with $S$, each ancilla is still in state $\eta$ and is fully uncorrelated with $S$ and all the other ancillas.

The collision between $S$ and the $n$th ancilla is described by the unitary operator $\hat {U}_{n}=e^{-i\left(\hat H_{S}+\hat{V}_{n}\right) \Delta t }$ ($\hbar\ug1$ throughout), where $\Delta t $ is the collision time, $\hat H_S$ the free Hamiltonian of $S$ and $\hat{V}_{n}$ stands for the coupling Hamiltonian between $S$ and the involved ancilla. A more general CM formulation can additionally feature a free Hamiltonian of the ancillas, which here is not necessary to consider.

After $n$ collisions, the joint system is given by $\sigma_n=\hat U_{n} \cdots\hat{ U}_{1}\,\sigma_{0}\,\hat U_{1}^\dag\cdots\hat{ U}_{n}^\dag$. This, alongside the crucial hypothesis that the initial state features no correlations, yields that the current state of $S$, $\rho_{n}={\rm Tr}_{B} \{\sigma_n\}$, depends on the state at the previous step as
\begin{eqnarray}
\rho_{n}=
{\rm Tr}_{n}\!\left\{\hat {U}_{ n }\left(\rho_{n-1}\,\eta\right)\,\hat {U}_{ n }^\dag\right\}={\cal E}[\rho_{n-1}]\,,\label{rhon1}
\end{eqnarray} 
where ${\rm Tr}_{n}$ is the partial trace over the $n$th ancilla and, importantly, $\rho'={\cal E}[\rho]$ defines a completely positive and trace-preserving (CPT) quantum map \cite{books1,books2} on the open system $S$.
By iteration, this yields that $\rho_n={\cal E}^n[\rho_{0}]$, showing that the evolution of $S$ occurs through repeated applications of the collision CPT map $\cal E$ on the initial state $\rho_0$. This fact, which is equivalent to \eq\eqref{rhon1} and can be regarded as a discrete version of the semigroup property \cite{books1,books2}, shows that the open dynamics of $S$ is fully {\it Markovian}: the knowledge of the system's state at the current step $n$ is enough to fully determine the evolution at all next steps $m>n$, regardless of evolution at previous steps $m<n$. 

The Markovian nature of the discrete dynamics discussed thus far supports the expectation that, passing to the continuous limit such that $t=n \Delta t$ becomes a continuous time, \eq\eqref{rhon1} gives rise to a Lindblad master equation. Conditioned to the requirement that this limit exists (see \eg\rref\cite{altamirano2017a}), this is indeed the case.

\section{Colored-noise bosonic reservoir}\label{bos-rev}

Assume now to have a generic system $S$ with free Hamiltonian $\hat H_S$ coupled to a continuum of bosonic modes (bosonic bath).
The free Hamiltonian of the bath reads
\begin{equation}
\hat H_f=\int\! {\rm d}\omega\,\omega\,\hat a^\dag(\omega)a^\dag(\omega)\,,\label{Hf}
\end{equation}
where $\hat a(\omega)$ [$\hat a^\dag(\omega)$] annihilates (creates) a photon of frequency $\omega$ and with the integral
running over the entire real axis (similarly for all the integrals appearing henceforth). Normal-mode ladder
operators fulfill the commutation rules $[\hat a(\omega), \hat a^\dag(\omega')]=\delta(\omega{-}\omega')$
and $[\hat a(\omega), \hat a(\omega')]=[\hat a^\dag (\omega), \hat a^\dag (\omega')]=0$.
The coupling between $S$ and the field is described by the interaction Hamiltonian 
\begin{equation}
\hat V=\sqrt{\frac{\gamma }{2\pi}}\int\! {\rm d}\omega\,\left(F(\omega)\, \hat b\, \hat a^\dag(\omega)+{\rm H.c.}\right) \,,\label{Vq0}
\end{equation}
where $\hat b$ and $\hat b^\dag$ are operators on $S$ and $\gamma$ has the dimensions of a rate. The dimensionless (generally complex) function $F(\omega)$ describes a  ``colored" coupling, thus generalizing the standard white-coupling case which is retrieved for $F(\omega)=1$. Function $F(\omega)$ is assumed to be ``well-behaved". We note that the same microscopic model was considered in some non-Markovian extensions of the usual input-output formalism \cite{diosi,jacobs}.

In the interaction picture with respect to $\hat H_0=\hat H_f$, the joint state of $S$ and the field 
evolves as
\begin{equation}
\frac{{\rm d}\sigma}{{\rm d}t}=-i [\hat H_S+ \hat V(t),\sigma]\label{dsigma1}
\end{equation}
with 
\begin{equation}
\hat V(t)=\sqrt{\frac{\gamma }{2\pi}}\int\! {\rm d}\omega\,\left(F(\omega) \hat b\, \hat a^\dag(\omega)e^{i\omega t}+{\rm H.c.}\right)\,.\label{Vt}
\end{equation}

\section{Time modes} \label{tm}

Since $\omega$ takes on values over the entire real axis, one can define {\it time} bosonic modes (or input modes in the language of input-output formalism \cite{gardiner}) through the Fourier transform of the field normal modes $\hat a(\omega)$ as
\begin{equation}
\hat a_{\rm in}(t)=\frac{1}{\sqrt{2\pi}}\int\! {\rm d}\omega\,\hat{a}(\omega)e^{-i\omega t}\,,\label{input}
\end{equation}
whose inverse formula reads
\begin{equation}
\hat a(\omega)=\frac{1}{\sqrt{2\pi}}\int\! {\rm d}t \,\hat{a}_{\rm in}(t)e^{i\omega t}\,.\label{input2}
\end{equation}
Time modes \eqref{input} fulfill bosonic commutation rules
\begin{equation}
[\hat a_{\rm in}(t), \hat a^\dag_{\rm{in}t}(t')]=\delta(t-t')\,,\label{comm}
\end{equation}
while $[\hat a_{\rm in}(t), \hat a_{\rm{in}t}(t')]=[\hat a^\dag_{\rm in}(t), \hat a^\dag_{\rm{in}t}(t')]=0$. 
Although they are not normal modes, time modes \eqref{input} are an alternative way to represent the field that is typically advantageous in many problems.

We next express the coupling Hamiltonian in the interaction picture, \eq\eqref{Vt}, in terms of time modes, obtaining
\begin{equation}
\hat V(t)=\sqrt{\gamma}\,\,\hat b\, \int\! {\rm d}t' \,{\cal F}(t-t')\,a^\dag_{\rm in}(t')+{\rm H.c.}\,\,,\label{Vt2}
\end{equation}
where the $\cal F$-function is defined as
\begin{eqnarray}
{\cal F}(t-t')=\frac{1}{2\pi}\int\! {\rm d}\omega\,F(\omega)e^{i\omega (t-t')}\,,\label{w-def}
\end{eqnarray}
thus representing the Fourier transform of the coupling function $F(\omega)$.
In the special case of white coupling, that is $F(\omega)=1$, we get ${\cal F}(t-t')=\delta(t-t')$ in a way that, at any given time $t$, $S$ couples only to a {\it single }time mode $\hat a_{\rm in}(t)$ (local coupling). This is a common situation in quantum optics, which upon suitable time discretization can be shown to lead to a CM where each ancilla collides with $S$ only {\it once} \cite{CMQO,fischerJPC}. \eq\eqref{Vt2} shows that, in the more general case of {colored} couplings, the interaction between $S$ and the bath is {\it non-local} in the time-mode representation. This gives rise to a CM that generally features {\it multiple} collisions between $S$ and each given ancilla, which is shown in the next section.

\section{Non-Markovian collision model}\label{timedis}

We next discretize time as $t_n=n\Delta t$ with $\Delta t$ the time step and $n$ an integer number. In each time interval, we define the field operator
\begin{align}
{\hat \alpha}_n=\frac{1}{\sqrt{\Delta t}}\int_{t_{n-1}}^{t_n}\!\!{\rm d}t'  \,\,{\hat a}_{\rm in}(t')\label{alphan}\,.
\end{align}
Due to \eq\eqref{comm}, operators $\{\hat \alpha_n\}$ fulfill bosonic commutation rules $[\hat \alpha_n,\hat \alpha^\dag_m]=\delta_{nm}$ and 
$[\hat \alpha_n,\hat \alpha_m]=[\hat \alpha_n^\dag,\hat \alpha^\dag_m]=0$. To work out the discrete version of \eqref{Vt2}, in each interval 
we replace $\hat V(t)$ with its coarse-grained approximation $\hat V_n$ as
\begin{equation}
\hat V(t)\simeq \hat V_n\ ~~~~{\rm for}\,\,\,t\in[t_{n-1},t_n], \label{coarse}
\end{equation}
where
\begin{align}
\hat V_n =\frac{1}{\Delta t}\int_{t_{n-1}}^{t_n}{\rm d}s\,\hat V(s)\label{Vt3-1}\,.
\end{align}

Next, in \eq\eqref{Vt3-1} we expand $\hat V(s)$ as [\cf\eq\eqref{Vt2}]
\begin{align}
\hat V(s)=\sqrt{\gamma}~\hat{b}~ \sum_m \int_{t_{m-1}}^{t_m}\! {\rm d}t' \,\,{\cal F}(s-t')\hat a_{\rm in}^\dag(t')\,,\label{Vs}
\end{align}
By plugging \eqref{Vs} into \eqref{Vt3-1},
$\hat V_n$ can be arranged in the form
\begin{align}
\hat V_n=\frac{\sqrt{\gamma}}{\Delta t}\,\,\hat b\, \sum_m \int_{t_{m-1}}^{t_m}\! {\rm d}t' \,\left(\int_{t_{n-1}}^{t_n}\!{\rm d}s \,\,{\cal F}(s-t')\right)a_{\rm in}^\dag(t')+{\rm H.c.}\label{Vt3}\,
\end{align}
Here $t_m=m\Delta t$, where $m$ is an integer number, is the discretized version of $t'$ (while $t_n=n\Delta t$ is the discretized version of $t$).
In the integral over $t'$ in \eq\eqref{Vt3}, note that the integrand features the input operator $\hat a_{\rm in}(t')$ weighted by a kernel $t'$-function (between big brackets) that depends on ${\cal F}$. For $\Delta t$ short enough \cite{nota-dt}, the kernel function can be replaced by its mean value in the interval $t'\in [t_{m-1},t_m]$, which we call $W_{nm}$, as \cite{nota-kernel}
\begin{align}
\int_{t_{n-1}}^{t_n}\!{\rm d}s \,\,{\cal F}(s-t')\simeq\frac{\int_{t_{m-1}}^{t_m}\!{\rm d}t'\,\,\left(\int_{t_{n-1}}^{t_n}\!{\rm d}s \,\,{\cal F}(s-t')\right)}{\Delta t}=W_{nm}\,\,.\label{W-def}
\end{align}
Quantity $W_{nm}$ can now be factored out of the integral over $t'$ in \eq\eqref{Vt3}, which yields
\begin{align}
\hat V_n=\frac{\sqrt{\gamma}}{\Delta t}\,\hat b\,\sum_m W_{nm}\, \int_{t_{m-1}}^{t_m}\! {\rm d}t' \,a^\dag_{\rm in}(t')+{\rm H.c.}\,\label{Vt4}
\end{align}
Recalling now definition \eqref{alphan} we get that for $\Delta t$ short enough
\begin{align}
\hat V_n=\,\hat b\,\sum_m g_{nm} \,\hat \alpha_m^\dag+{\rm H.c.}\label{Vt4}\,,
\end{align}
where the coupling strengths $g_{nm}$ are defined as \cite{nota-spirit}
\begin{align}
g_{nm}=\sqrt{\frac{\gamma}{\Delta t}}\,W_{nm}\,\,.\label{gnm}
\end{align}
In each time interval $t\in[t_{n-1},t_n]$, the joint system evolves under the coarse-grained Hamiltonian $\hat H_S+\hat V_n$. The corresponding time evolution operator in the same interval, i.e., during the $n$th collision, thus reads
\begin{equation}
\hat {U}_{n}\simeq \hat{\mathbb I}-i (\hat{H}_{S}+{\hat V_{n} })\Delta t -\tfrac{{\hat V}_{n}^2}{2}\Delta t ^2\,,\label{USn2}
\end{equation} 
where we retained terms up to the second order in $\Delta t$.

Just like the white-noise case \cite{CMQO}, one can thus define a CM such that modes \eqref{alphan} embody the ancillas. Yet, at the $n$th step, a collision occurs in which the system non-locally interacts with {\it all} of the ancillas with coupling rates $g_{nm}$ [\cf\eq\eqref{Vt4}], instead of a single one as in the white-noise case. A sketch of the CM dynamics is shown in \fig1(b), which can be compared with the memoryless case (white-noise bath) in \fig1(a).
Note that, unlike standard notation used for memoryless CMs \cite{CMQO}, here subscript $n$ in $\hat U_n$ and $\hat V_n$ solely labels the time step since these operators generally act on $S$ and {all} of the ancillas (not just $S$ and the $n$th subunit). 

Clearly, the nature of collisions [see \fig1(b)] is such that each given ancilla generally collides with $S$ at different steps, not just one as in standard memoryless CMs [see \fig1(a)]. As a major consequence, at each step $S$ in particular collides with ancillas it has already interacted and established correlations with at previous steps: applying unitary \eqref{USn2} on the current system-bath state and tracing off ancillas does {\it not} return a completely positive map, preventing ``CP divisibility" \cite{review1,review2} and thus a Lindblad master equation (even a fully time-dependent one) to hold.

\subsection{White coupling}\label{white}

As anticipated in Section \ref{bos-rev}, the standard white-noise model used in quantum optics \cite{gardiner} is retrieved for $F(\omega)=1$. \eq\eqref{w-def} then reduces to a delta function, ${\cal F}(t-t')=\delta (t-t')$,  leading to [\cf\eq\eqref{W-def}] $W_{nm}=\delta_{nm}$. Thereby, in virtue of \eqs\eqref{Vt4} and \eqref{gnm}, the resulting CM reduces to the basic CM in \fig1(a) where each ancilla collides with the system only once. Such a CM entails Markovian dynamics and Lindblad ME \cite{CMQO} provided that ancillas are initially uncorrelated. The last condition is due to the fact that, even if double collisions do not occur and ancillas are noninteracting, initially correlated ancillas generally give rise to non-Markovian dynamics \cite{filippov} (see also Sect.2.2 in \cite{CMQO}), an important instance being single-photon states \cite{Braciola,Dabrowska,NJPNM}.


\section{An atom in  a semi-infinite waveguide}\label{mirror}

Consider a two-level atom $S$ coupled under the rotating-wave approximation at $x=x_0$ to a semi-infinite waveguide lying along the positive $x$-semi axis (this can be seen as an infinite waveguide with a perfect mirror at $x=0$). The setup is sketched in \fig2(a). The waveguide has linear dispersion law $\omega=v k$ with $v$ the photon group velocity. The atom's ground and excited states are denoted by $|g\rangle$ and $|e\rangle$, respectively, their energy separation being $\omega_0=v k_0$. It is easily shown \cite{NJPNM,sore} that upon a waveguide ``unfolding" at $x=0$ the system maps into an atom coupled at {\it two} points, $x=\pm x_0$, to a {\it chiral} waveguide extending over the entire $x$-axis [see \fig2(b)]. The corresponding Hamiltonian in $k$-space reads
 \begin{align}
 \hat H\,=\,&\omega_0|e\rangle\langle e|+\int \!d k \,v k \,\hat a_k^\dag \hat a_k\nonumber\\
 &+\sqrt{\frac{\gamma v}{2\pi}}\int \!d k \, \left(\left(e^{i (k_0+k) x_0}-e^{-i( k_0+k) x_0}\right)|g\rangle\langle e |\hat a_k^\dag+{\rm H.c.}\right)\,,\nonumber
 \end{align} 
 where $\gamma$ is half the decay rate of the atom in the absence of the mirror. 
\begin{figure}
	\includegraphics[width=\textwidth]{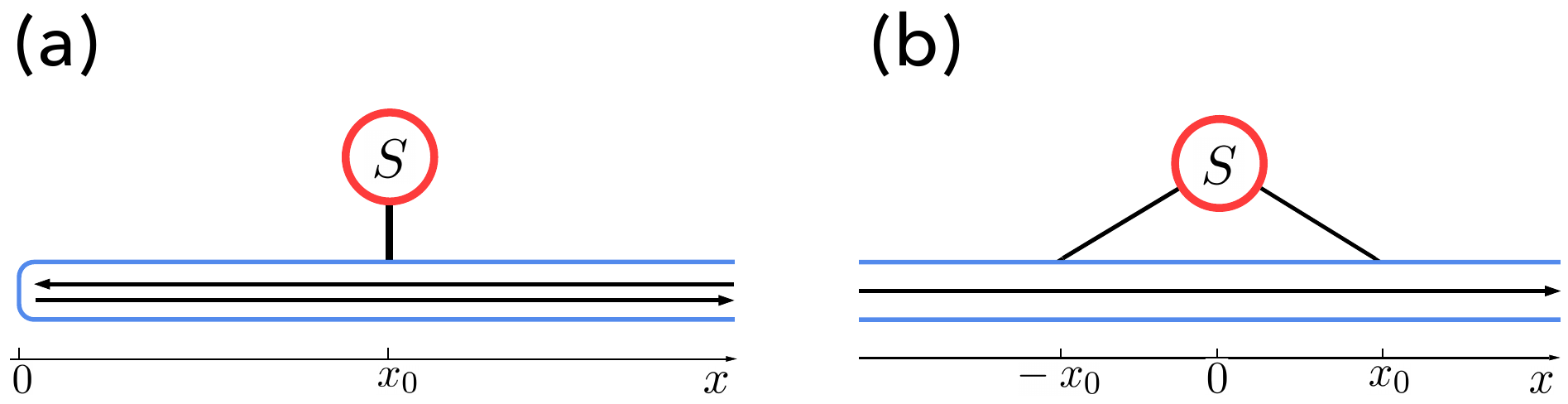}
	\caption{(a) An atom $S$ coupled at $x=x_0$ to a semi-infinite waveguide. The waveguide's end at $x=0$ can be seen as a perfect mirror. (b) Atom coupled to a chiral waveguide at two points, $x=\pm x_0$. The dynamics of system (a) can be effectively mapped into that of setup (b).} 
\end{figure}

Making the variable change $x\rightarrow x-x_0$ (in a way that the atom now couples to the waveguide at $x=0$ and $x=-2x_0$) and passing to the frequency domain $\omega=v k$, we get a system-bath Hamiltonian of the same type as the one assumed in Section \ref{bos-rev} [\cf \eq\eqref{Vq0}] with $\hat b=|g\rangle\langle e|$ and
\begin{align}
F(\omega)=1-e^{-i \phi}e^{-i \omega \tau}\,.\label{Fw}
\end{align}
Here, $\tau=2 x_0/v$ ({\it time delay}) is the time taken by a photon resonant with the atom to complete a round trip between the atom and mirror, while $\phi=2k_0 x_0$ is the phase shift acquired along the same path \cite{milonni1,milonni2,dorner,tufa2}.

Hence, the presence of the mirror ``colours" the system-bath coupling function. Specifically, this has sinusoidal behavior with ``period" $\sim1/\tau$, thus the longer the time delay the more structured is $F(\omega)$. 

Using \eq\eqref{w-def}, the ${\cal F}$-function corresponding to \eq\eqref{Fw} reads
\begin{align}
{\cal F}(t-t')=\delta(t-t')-e^{-i\phi}\delta(t-\tau-t')\,,\label{Ft2}
\end{align}
which shows that, compared to white coupling (Subsection \ref{white}), an extra delta function arises that is centered at $t-\tau$. Thus ${\cal F}(t-t')$ is non-zero only at $t'=t$ and $t'=t-\tau$. This ``bi-local" behavior stems from the fact that there is a precise instant at which a photon emitted from the atom returns to it after bouncing back from the mirror. 

\begin{figure}\centering
	\includegraphics[width=6cm]{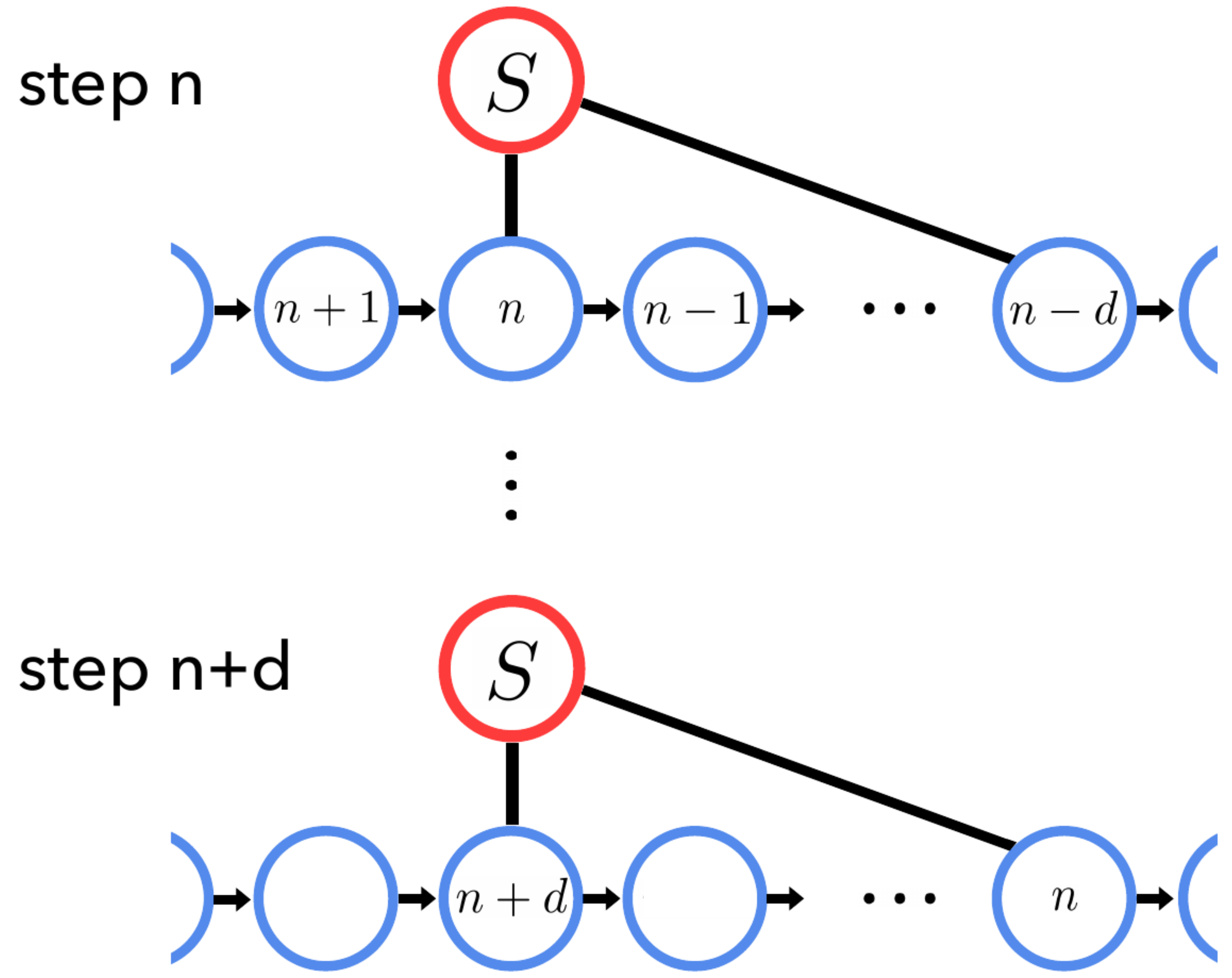}
	\caption{Non-Markovian collision model for an atom emitting in front of a mirror in the case of non-negligible retardation times corresponding to coupling function \eqref{Fw}. At step $n$, $S$ collides with {\it two}  ancillas, $n$ and $n-d$, where $d$ is the time delay in units of time step $\Delta t$. Hence, each given ancilla $n$ collides with $S$ twice, the first time at step $n$ and then at step $n+d$.} \label{fig1}
\end{figure}
Next, in order to work out the CM in this case, we plug \eqref{Ft2} into \eq\eqref{W-def} obtaining
\begin{align}
W_{nm}&=\delta_{nm}-e^{-i\phi}\delta_{n-d,m}\label{W-def-2}
\end{align}
where integer $d=\left[\frac{\tau}{\Delta t}\right]$ is in practice the time delay in units of $\Delta t$.  Combined with \eqs\eqref{Vt4} and \eqref{gnm}, this yields that at step $n$ of the CM dynamics the atom collides with ancillas $m=n$ and $m=n-d$, the corresponding coupling Hamiltonian being
\begin{align}
\hat V_n=\sqrt{\tfrac{\gamma}{\Delta t}}\,\,|g\rangle\langle e|\left(\hat \alpha_n^\dag-e^{-i\phi}\hat \alpha_{n-d}^\dag\right)+{\rm H.c.}\label{Vt5}\,.
\end{align}
A pictorial sketch of this CM is shown in \fig3. Despite the collisions pattern is considerably simpler than the general case in \fig1(b), the dynamics is generally non-Markovian \cite{NJPNM,tufa} and hard to tackle. This CM was first investigated in \cite{grimsmo2015,pichler}, which proposed methods to attack the problem when the atom is driven by a classical field. 

Nevertheless, for spontaneous emission (field initially in the vacuum state) the atom's excited-state amplitude $\varepsilon$ in the continuous-time limit ($\Delta t\rightarrow 0$) can be shown to obey the exact time-non-local differential equation \cite{milonni1,milonni2,dorner,tufa2}
\begin{align}
\dot \varepsilon(t)=-(i\omega_0+\gamma)  \varepsilon(t)+\gamma e^{i \phi}\varepsilon(t-\tau)\theta(t-\tau)\,,\label{DDE-c}
\end{align}
which can be analytically solved. The resulting non-Markovian decay in the long-delay regime has been recently measured in a closely related setup \cite{delsing}.
Using standard methods \cite{books1,books2} (not based on CMs), \eq\eqref{DDE-c} can be worked out without big efforts from the spectral density \cite{tufa} this being essentially the squared modulus of \eqref{Fw}.
However, for illustrative purposes, in the next section we present a proof of \eq\eqref{DDE-c} entirely formulated in terms of the CM in \fig3.

\section{Collision-model derivation of \eq\eqref{DDE-c}}\label{DDE}

The initial state of $S$ and ancillas reads $\ket{\Psi^{(0)}}=\ket{e}_S\otimes_m\!\ket{0}_m$, with $|0\rangle_m$ and $|1\rangle_m$ the ancilla's vacuum and single-photon Fock states, respectively. Based on \eq\eqref{Vt5}, each collision conserves the total number of excitations $\hat N=|e\rangle\langle e|+\sum_m \hat\alpha_m^\dag\hat\alpha_m$. Thereby, since $N=1$ for the initial state, the dynamics takes place entirely within the single-excitation sector and the joint state at any step $n$ has the form
\begin{align}
\ket{\Psi^{(n)}} = \varepsilon^{(n)}\ket{1_S} + \sum_m c^{(n)}_m  \ket{1_m}\,.
\label{st}
\end{align}
Here, we used a compact notation such that $\ket{1_S}=\ket{e}\otimes_m\!\ket{0}_m$ (one excitation on the atom), while in $\ket{1_m}$ the atom is unexcited and all ancillas in the vacuum state but the $m$th one which is in $|1\rangle_m$. Superscript ``$(n)$" refers to the time step, while subscript ``$m$" labels the ancillas. Note that ancillas behave as effective qubits. 
In terms of the excitation amplitudes in \eq\eqref{st}, the initial state $|\Psi^{(0)}\rangle$ reads $\varepsilon^{(0)}=1$, $c_m^{(n)}=0$ for any $m$. For convenience, we will assume that excitation amplitudes are defined also for negative values of the step index $n$: $\varepsilon^{(n\le 0)}=1$, $c_m^{(n\le 0)}=0$.

At each step the joint state transforms as
$\ket{\Psi^{(n+1)}} =  \hat{U}_n \ket{\Psi^{(n)}},$
where $\hat U_n$ is the unitary describing the evolution in each collision [recall \eq\eqref{USn2}]. 
When $\hat U_n$ is calculated using \eq\eqref{Vt5}, applied to \eqref{st} and the resulting state projected to $|1_S\rangle$, we get
\begin{align}
&
\varepsilon^{(n+1)}= \varepsilon^{(n)} -(i\omega_0+\gamma) \Delta t \,\varepsilon^{(n)}-i \sqrt{\gamma\Delta t}  \left( c_{n}^{(n)} - e^{i \phi} c_{n-d}^{(n)} \right) \,. 
\label{eps}
\end{align}
We observe now that $c_m^{(n)}=0$ for any $m\ge n$ since the corresponding ancillas have not yet interacted with $S$ and thus are still unexcited (ancillas on the left of $S$ in \fig3 including the one in front of it). Thus, in particular, $c_n^{(n)}$ can be set to zero in \eq\eqref{eps}, which reduces to 
\begin{align}
&
\varepsilon^{(n+1)}-\varepsilon^{(n)} = -(i\omega_0+\gamma) \Delta t \,\varepsilon^{(n)}+i \sqrt{\gamma\Delta t}  \,e^{i \phi} c_{n-d}^{(n)}  \,. 
\label{eps2}
\end{align}
We will first consider the case $n\ge d$, then $0\le n<d$. 

In \eq\eqref{eps2} we need to eliminate $c_{n-d}^{(n)}$. To this aim, we note [see \fig3] that ancilla $n-d$ collides with $S$ the first time at step $n-d$ and then at step $n$. It follows that the corresponding amplitude at step $n-d+1$ no longer changes until step $n$
\begin{align}
c_{n-d}^{(n)} =c_{n-d}^{(n-1)}=\dots= c_{n-d}^{(n-d+1)}\,.\label{chain}
\end{align}
Amplitude $c_{n-d}^{(n-d+1)}$ can be worked out similarly to \eq\eqref{eps} by applying the collision unitary to $ \ket{\Psi^{(n-d+1)}}$ and projecting next to $|1_{n-d}\rangle$. This yields
\begin{align}
c_{n-d}^{(n-d+1)} =   -i\sqrt{\gamma\Delta t} \,\varepsilon^{(n-d)} +\tfrac{1}{2}\gamma \Delta t \, 
e^{i \phi} c_{n-2d}^{(n-d)}\label{cnd2}\,.
\end{align}

When this is replaced in $c_{n-d}^{(n)}$, \eq\eqref{eps2} becomes  
\begin{align}
\varepsilon^{(n+1)}-\varepsilon^{(n)}=-(i\omega_0+\gamma) \Delta t \varepsilon^{(n)}+\gamma \Delta t e^{i \phi}\varepsilon^{(n-d)}\nonumber\,,
\end{align}
where we neglected terms of order higher than $\gamma \Delta t$. Setting $\Delta \varepsilon_n=\varepsilon^{(n+1)}-\varepsilon^{(n)}$, this is turned into the finite-difference equation
\begin{align}
\frac{\Delta \varepsilon_n}{\Delta t}=-(i\omega_0+\gamma) \varepsilon^{(n)}+\gamma  e^{i \phi}\varepsilon^{(n-d)}\,.\label{DDE-fin}
\end{align}
Taking the continuous-time limit $\Delta t \rightarrow 0$, such that $\varepsilon_n\rightarrow \varepsilon (t)$ and $\varepsilon^{(n-d)}\rightarrow \varepsilon(t-\tau)$, \eq\eqref{DDE-fin} reduces to \eq\eqref{DDE-c} for $t>\tau$. 

To complete the proof, we note that, for $0\le n< d$, \eq\eqref{eps2} misses the last term due to the initial conditions. This immediately yields \eq\eqref{DDE-fin} but without the term $\propto \varepsilon^{(n-d)}$, hence in the continuous-time limit we end up with \eq\eqref{DDE-c} for $0\le t< \tau$. 
\\
\\
The above derivation was intended to provide an analytically solvable instance of a non-Markovian collisional dynamics, which can be contrasted with usual memoryless CM dynamics as well as with non-Markovian ones that yet can be made Markovian by embedding $S$ into a larger system \cite{lorenzo2016,luoma2016,lorenzo2017,steve}.  
In the latter respect, note that a form of embedding for the CM in \fig3 was shown through an elegant diagrammatic technique in \rref\cite{grimsmo2015}, although the Hilbert-space dimension of the extended Markovian system scales as $\delta^\nu$ with $\delta$ the dimension of $S$  and $\nu=[t/\tau]$.

\section{Conclusions}\label{concl}

Starting from the known way to build up a CM for a system coupled to a white-noise bosonic reservoir, we discussed how the CM construction can be extended to the case of colored system-bath coupling. While fictitious ancillas are defined identically to the white-noise case, having a structured coupling function yields a collisional picture where, in the general case, the system interacts with all the bath ancillas at each step. This is an intrinsically non-Markovian CM in that each collision is not described by a CPT map on the system. We illustrated this for an atom emitting in front of a mirror in the regime of non-negligible delay times, in which case the system collides with two ancillas at each step. Finally, we used this non-Markovian CM to retrieve the known (time-non-local) delay differential equation governing the atomic decay.

We point out that the regime of non-negligible time delays (typically neglected in traditional quantum optics) is currently the focus of a growing literature \cite{grimsmo2015,pichler,NJPNM,Laakso, Ramos, Tabak, Guimond,PiCHoiZol,Guo,Chalabi,Nemet,FangCom,Calajo}
. First experiments are underway in setups featuring slowly propagating fields (so as to lengthen time delays). For instance, the non-Markovian decay predicted by \eq\eqref{DDE-c} has been recently measured by coupling a superconducting qubit to surface acoustic waves at two distant points \cite{delsing}.

\section*{Acknowledgements}

Fruitful discussions with Susana Huelga, Kimmo Luoma, Gonzalo Manzano, Salvatore Lorenzo and Tommaso Tufarelli are gratefully acknowledged.


\begin{thebibliography}{8}
\bibitem{nota-kernel} The integral on the left-hand side of \eq\eqref{W-def} is here viewed as a function of $t'$, which can be approximated by its coarse-grained version in the corresponding interval $[t_{m-1},t_{m}]$ [as done in \eqs\eqref{coarse} and \eqref{Vt3-1}].
\bibitem{nota-spirit} In virtue of \eq\eqref{alphan}, the basic spirit of the calculation leading to \eq\eqref{Vt4} is to formally end up with integrals featuring only $\hat a_{\rm in}(t)$ in the integrand.
\end{thebibliography}
\end{document}